\documentstyle[amssymb,aps]{revtex}

\begin{document}
\author{J. L. Cohen, B. Dubetsky, and P. R. Berman}
\address{Department of Physics, University of Michigan, Ann Arbor, MI 48109-1120}
\title{Quasiperiodic Atom Optics, Focusing, and Wave packet Rephasing}
\date{September 23, 1998}
\maketitle

\begin{abstract}
We propose a laser field configuration which acts as a quasiperiodic atom
optical diffraction grating. Analytical and computational results for the
atomic center-of-mass wavefunction after the grating reveal a quasiperiodic
density pattern, a semiclassical focusing effect, and a quasiperiodic
self-imaging of the atomic wavefunction analogous to a Talbot effect.
\end{abstract}


\label{sec:1}

Increased attention has been given to periodic atom optical systems and to
drawing parallels between solid state physics and the corresponding atom
optics in an easily-controlled environment\cite{i}. In the transient
interaction regime, periodic atom optical elements have been used for
spatial-\cite{1} and time-domain\cite{2a,2} interferometry, atom focusing
and lithography \cite{3}, and the Fresnel self-imaging (Talbot effect\cite{7}%
) of an atomic wave packet\cite{7aaa}. In addition, optical lattices\cite{7a}
have been used to simulate solid state effects using atomic de Broglie
waves. These atoms exhibit quantized motion\cite{7aa}, extended/localized
state transitions\cite{8}, and Bloch oscillations\cite{10}. To further the
connections with condensed matter and classical optics, the extension of
atom optical experiments to quasiperiodic systems seems natural \cite
{11a,11b,14a}.

Recently, Guidoni and coworkers \cite{12} trapped cesium atoms into a
quasiperiodic optical lattice formed by a three-dimensional laser
configuration with incommensurate spatial components of intensity. Qian Niu
and coworkers have been analyzing such optical lattices in one- and
two-dimensions to understand their eigenstructure\cite{13}. We propose to
extend atom lithography and interferometry experiments to create atomic
beams or cold trapped atoms with quasiperiodic center-of-mass wave
functions. We present analytical and computational results for the atomic
wave function and density after interaction with a one-dimensional,
quasiperiodic atom optical diffraction grating. Momentum distributions
reveal the quasiperiodic nature of the wave packets. Furthermore, we show
the possibility for wave packet revivals, essentially a quasiperiodic Talbot
effect, where the initial atomic wave function can be (nearly) recovered.

A schematic of an experiment, similar to typical atomic focusing and Talbot
arrangements, is shown in Fig. 1\cite{3,7aaa}. Two pairs of off-resonant
laser beams of width $w$ intersect at a point along a monoenergetic,
transversely-cooled atomic beam with velocity $v_{z}$ propagating in the $z$%
-direction (or alternatively in the $y$-direction). Assuming the laser beam
pairs are detuned from one another, we can ignore the cross-terms in the
intensity pattern. The atoms are modeled as two-level center-of-mass plane
waves with upper-state lifetimes of $\Gamma ^{-1}$. Hence, this optical
configuration forms a light-shift potential for the atomic ground state, 
\begin{equation}
V(x,z)=V_{1}(z)\cos 2kx+V_{2}(z)\cos \sqrt{2}kx,  \label{1}
\end{equation}
where $V_{n}(z)\simeq \hbar \left| \Omega _{n}(z)\right| ^{2}/8\Delta _{n}$
for Rabi frequencies $\Omega _{n}(z)$ and atom-field detunings $\Delta _{n}$
of the laser beam pairs $n=1,2$. Spontaneous emission during the atom-field
interaction is ignored by assuming that $\left| \Omega _{n}(z)\right|
^{2}\Gamma w/4v_{z}\Delta _{n}^{2}\ll 1$. The quasiperiodicity arises from
the incommensurate wave vectors of the optical potential, and phase
stability between the laser fields is essential for the potential's
integrity.

The atomic motion is described by an effective Schr\"{o}dinger equation for
the transverse wave function $\phi (x,t)$ in the atomic rest frame, $%
z=v_{z}t.$ Assuming $1/2Mv_{z}^{2}\gg \left\langle
1/2Mv_{x}^{2}+1/2Mv_{y}^{2}+V(x,z)\right\rangle $, this equation reads\cite
{14c} 
\begin{equation}
i\hbar \frac{\partial \phi }{\partial t}=\left[ \frac{p_{x}^{2}}{2M}+V(x,t)%
\right] \phi .  \label{2}
\end{equation}
The potential in the interaction region appears as a pulse of duration $\tau
=w/v_{z}$ in the atomic rest frame. The pulse shape is determined by the
transverse laser profiles.

The discussion here is restricted to the Raman-Nath regime for square
pulses, $V(x,t)=V(x)$ for $-\tau \leq t\leq 0$ and $V(x,t)=0$ for all other
times, where 
\begin{equation}
V(x)=V_{1}\cos 2kx+V_{2}\cos \sqrt{2}kx\text{.}  \label{3}
\end{equation}
In the Raman-Nath approximation one assumes that $\omega _{k}\left|
V(x)\right| \tau ^{2}/\hbar \ll 1,$ where $\omega _{k}=2\hbar k^{2}/M$ is a
two-photon recoil frequency; the kinetic energy is ignored during the
interaction, allowing an immediate integration of Eq. (\ref{2}) using $\phi
(x,-\tau )=1$ and Eq. (\ref{3})\cite{14cc}. We can write $\phi (x,0)=\exp %
\left[ -iV(x)\tau /\hbar \right] $ or 
\begin{equation}
\phi (x,0)=\exp \left[ i(A_{1}\cos 2kx+A_{2}\cos \sqrt{2}kx\right] \text{,}
\label{4}
\end{equation}
where the pulse area $A_{n}=-V_{n}\tau /\hbar $. The standing-wave light
fields acts as a quasiperiodic atomic phase grating.

To follow the evolution of the wave packet after the interaction, we can
expand the exponentials in Eq. (\ref{4}) in a plane wave representation. The
resulting, initial wave function for the time-dependent Schr\"{o}dinger
equation is a superposition of free particle eigenstates $\exp
[ip_{m,n}x/\hbar ]$ with energies $E_{m,n}=p_{m,n}^{2}/2M$, giving the
result 
\begin{equation}
\phi (x,t>0)=\sum_{m,n=-\infty }^{\infty
}i^{m+n}J_{m}(A_{1})J_{n}(A_{2})\exp \left[ i2kx(m+\frac{n}{\sqrt{2}})\right]
\exp \left[ -i\varphi _{m,n}(\omega _{k}t)\right] ,  \label{5}
\end{equation}
where 
\begin{equation}
\varphi _{m,n}(\omega _{k}t)=E_{m,n}t/\hbar =(m+\frac{n}{\sqrt{2}}%
)^{2}\omega _{k}t  \label{6}
\end{equation}
is the phase of the momentum component 
\begin{equation}
p_{m,n}=2\hbar k(m+n/\sqrt{2})  \label{6'}
\end{equation}
and $J_{m}$ is a Bessel function of the first kind. The momentum space wave
function superposes a set of regularly spaced, but not periodic, components
which are integer combinations of momentum exchanges between the atom and
fields. By squaring Eq. (\ref{5}) and using a sum rule for Bessel function
products, the transverse atomic density, $\rho (x,t)=\phi ^{\ast }(x,t)\phi
(x,t)$, can be written as 
\begin{equation}
\rho (x,t)=\sum_{m,n}\rho _{m,n}(\omega _{k}t)\exp \left[ i2kx(m+\frac{n}{%
\sqrt{2}})\right] .  \label{6a}
\end{equation}
The density has time-dependent Fourier amplitudes 
\begin{eqnarray}
\rho _{m,n}(\omega _{k}t) &=&J_{m}\left( 2A_{1}\sin [(m+\frac{n}{\sqrt{2}}%
)\omega _{k}t]\right)  \nonumber \\
&&\times J_{n}\left( 2A_{2}\sin [(\frac{m}{\sqrt{2}}+\frac{n}{2})\omega
_{k}t]\right) \text{,}  \label{7}
\end{eqnarray}
creating a spatial pattern which evolves in time. Thus, the atomic density
is not only a quasiperiodic function of the coordinate $x$, but of the
coordinate $z=v_{z}t$ as well.

The density Fourier transform (DFT), 
\begin{equation}
\rho (q,t)=\int \frac{dx}{2\pi }\rho (x,t)e^{-iqx}=\sum_{m,n}\rho
_{m,n}(\omega _{k}t)\delta (q-p_{m,n}/\hbar )\text{,}  \label{7a}
\end{equation}
has peaks at $q=2k(m+n/\sqrt{2})$ by Eq. (\ref{6'}). When squared, $\rho
(q,t)$ gives the time-dependent structure factor of the atomic distribution.
In realistic experiments the delta-function lineshape of each spectral
component would be broadened according to the initial momentum distribution
of the transverse atomic beam. For a thermal velocity distribution with most
probable speed $u$, the replacement $\rho _{m,n}(\omega
_{k}t)\longrightarrow \rho _{m,n}(\omega _{k}t)\exp \left[ -\left(
p_{m,n}ut/2\hbar \right) ^{2}\right] $ in Eq. (\ref{7}) is sufficient to
account for Doppler dephasing.

To detect the density as a function of $t$, one can scatter a transient
probe off of the atoms at $t$ to record the time evolution of certain
Fourier components of the density. For example, a weak probe pulse with
duration $<(ku)^{-1}$ and wave vector ${\bf k}_{p}=-k\widehat{x}$
backscatters a field $E_{bs}$ proportional to $\rho _{1,0}(\omega _{k}t)$ in
the $+\widehat{x}$-direction: $E_{bs}$ $\sim J_{1}\left( 2A_{1}\sin [\omega
_{k}t]\right) J_{0}\left( 2A_{2}\sin [\omega _{k}t/\sqrt{2}]\right) \exp %
\left[ -\left( kut\right) ^{2}\right] $. This is a type of free induction
decay experiment to detect ground state population gratings\cite{14e}.

More importantly, either direct atomic deposition or lithography using the
atomic beam to impinge on a prepared substrate would reconstruct the atomic
density at a fixed time. Atomic lithography has advanced to the point where
atoms can carve nanostructures in materials such as silicon, silicon
dioxide, and gold\cite{14aa}. Such quasiperiodic surfaces could be used for
solid state surface and transport studies. The implications for quantum and
optical properties, including photon localization, may be profound owing to
the quasiperiodic boundary conditions for the electron or optical waves\cite
{14ee}.

We now examine two different phenomena in the pattern formed by the atoms (%
\ref{6a}), ignoring dephasing. Semiclassical (near-field) dynamics explain a
focusing effect, similar to that seen after periodic phase gratings\cite{3}.
In Fig. 2 the optical potential is shown for $A_{1}=5$ and $A_{2}=10$. Each
potential well acts as a lens which can focus atoms using the impulsive
(dipole) force, $F(x)=M\Delta v(x)/\tau =-\partial V(x)/\partial x$, where $%
\Delta v(x)$ is an impulsive velocity kick. To illustrate this effect, $V(x)$
is Taylor expanded around its minimum at $x=0$ to give the focusing force
near this point, $F(x)=M\Delta v(x)/\tau $ $\approx (4V_{1}+2V_{2})kx$.
Solving for $\Delta v(x)$ and setting $t_{f}=x/\Delta v(x),$ this
geometrical argument yields a focus at the time $\omega
_{k}t_{f}=(2A_{1}+A_{2})^{-1}$ that translates into a spatial distance $%
z_{f}=v_{z}t_{f}$. The ratio of pulse areas in Fig. 2, $A_{2}/A_{1}=2$, was
chosen so that each standing wave contributes an equal semiclassical force.

The density at this ''quasi''-focus with its peak at $x=0$ is also shown in
Fig. 2. Additional density peaks result from focusing by the shallower wells
which occur at the quasiperiods of the potential. For example, the peak at $%
2kx\approx 7\ast 2\pi \approx 5\sqrt{2}\ast 2\pi $ occurs near a potential
well where the incommensurate standing waves are nearly in phase. In
general, the irrational wave vector ratio, $\sqrt{2}$ in this case, can be
approximated as the ratio $a_{s}/b_{s}$, where $a_{s}$ and $b_{s}$ are
positive integers without common factors. Quasiperiods will then be defined
by $2kx\approx 2\pi ja_{s}\approx 2\pi jb_{s}\sqrt{2}$ for any integer $j$.
A converging sequence $G_{s}$ which approximates $\sqrt{2}$ is given in
Table 1 \cite{15}. 
\begin{eqnarray*}
&& 
\begin{tabular}{|c|c|c|c|c|c|}
\hline
$s$ & $2$ & $3$ & $4$ & $5$ & $6$ \\ \hline
$a_{s}$ & $3$ & $7$ & $17$ & $41$ & $99$ \\ \hline
$b_{s}$ & $2$ & $5$ & $12$ & $29$ & $70$ \\ \hline
$G_{s}=a_{s}/b_{s}$ & $1.5$ & $1.4$ & $1.4167$ & $1.4138$ & $1.4143$ \\ 
\hline
$\left| \sqrt{2}-G_{s}\right| $ & $8.6$e$-2$ & $1.4$e$-2$ & $2.5$e$-3$ & $%
4.2 $e$-4$ & $7.2$e$-5$ \\ \hline
$\omega _{k}t_{s}$ & $4\pi $ & $20\pi $ & $24\pi $ & $116\pi $ & $140\pi $
\\ \hline
\end{tabular}
\\
&&\text{Table 1. Sequence of approximations for }\sqrt{2}\text{ defines the }
\\
&&\text{spatial}\text{ }\text{quasiperiods, }2kx\approx 2\pi ja_{s}\text{,
and the }\text{quasi-Talbot } \\
&&\text{times }t_{s}\text{ in Eq. (\ref{8}).}
\end{eqnarray*}

Several peaks in Fig. 2 are labeled by their values of $(ja_{s},jb_{s})$.
For smaller differences between $ja_{s}\ $and $jb_{s}\sqrt{2}$, the
quasiperiodicity is more pronounced (i.e., the density peaks near $%
2kx\approx 2\pi ja_{s}$ approach the size of the peak at $x=0)$. Thus, the
density reflects the quasiperiodicity of the system.

The DFT contains the spectral information important for lithography or
scattering experiments. In Fig. 3 we plot the Fourier amplitudes $\left|
\rho _{m,n}(\omega _{k}t)\right| $ [Eq. (\ref{7})] at the wave vectors $%
q=2k(m+n/\sqrt{2})>0$ and at $t=t_{f}$ for $A_{1}=5$ and $A_{2}=10$ again:
this is the DFT of Fig. 2. The major peaks are labeled by $(m,n)$ to show
the variation of amplitudes. The inset of Fig. 3 has the same axes as the
main graph and shows the range of $q$ values with significant amplitudes.
This Fourier spectrum has qualitative scaling properties: for any Fourier
wave vector of the density $q_{0}$, a wave vector $q^{\prime }=2k(m^{\prime
}+n^{\prime }/\sqrt{2})$ can be found which is arbitrarily close to $q_{0}$,
even if the amplitude of that component is much less than one.

Atoms that propagate after interacting with periodic atom optical elements
exhibit a self-imaging of their wave function - Talbot effect - and return
to uniform density at times $\omega _{k}t=2\pi j$ for integer $j>0$\cite
{7,7aaa}. While an exact self-imaging which reproduces $\phi (x,0)$ [Eq. (%
\ref{4})] is impossible owing to the dispersion in Eq. (\ref{6}), the
quasiperiodic nature of the wave function can lead to a quasi-rephasing when 
$\varphi _{m,n}\simeq 2\pi j^{\prime }$ for some integer $j^{\prime }>0$.

The appropriate observation times for self-imaging will produce phases for
each momentum component which are nearly integer multiples of $2\pi $. For
odd values of $b_{s}$, the choice, $\omega _{k}t_{s}=4\pi b_{s}$, from Eq. (%
\ref{6}) gives the phases 
\begin{equation}
\varphi _{m,n}(4\pi b_{s})=2\pi (2b_{s}m^{2}+b_{s}n^{2}+2b_{s}\sqrt{2}mn),
\label{8}
\end{equation}
where we again refer to Table 1, and $b_{s}\sqrt{2}\simeq a_{s}$ by
construction. For even values of $b_{s},$ the rephasing occurs at $\omega
_{k}t_{s}=2\pi b_{s}$ from Eq. (\ref{8}). At these times the first two terms
in Eq. (\ref{8}) are integers for all $m,n$. Furthermore, the third term is
nearly an integer, as required.

The wave function phase $\theta (x,t)$ is defined by $\phi (x,t)=\left| \phi
(x,t)\right| \exp [i\theta (x,t)]$. For exact self-imaging, the wave
function should be the unitary exponential, Eq. (\ref{4}), which has density
equal to one and phase $\theta (x,0)=A_{1}\cos (2kx)+A_{2}\cos (\sqrt{2}kx)$%
. In Fig. 4 we plot $\theta (x,t_{s})$ for $A_{1}=$ $A_{2}=1$ and $s=2,4,$
and $6$. The self-imaging becomes more pronounced at longer times $t_{s}$,
corresponding to a better approximation of $\sqrt{2}$ by $G_{s}$. The
average values (denoted by the bar) and standard deviations (denoted by $%
\sigma $) of both $\rho (x,t_{s})$ and the phase difference, $\delta
_{s}=\theta (x,t_{s})-\theta (x,0)$, are shown in Table 2 for the cases of
Fig. 4. 
\begin{eqnarray*}
&& 
\begin{tabular}{|c|l|l|l|}
\hline
$s$ & $2$ & $4$ & $6$ \\ \hline
$\bar{\rho}(x,t_{s})$ & 1.019 & 1.00085 & 1.000024 \\ \hline
$\sigma (\rho (x,t_{s}))$ & 0.505 & 0.0170 & 0.000508 \\ \hline
$\overline{\delta _{s}}$ (rads) & 0.0324 & 0.00645 & 0.00106 \\ \hline
$\sigma (\delta _{s})$ (rads) & 0.417 & 0.0899 & 0.0156 \\ \hline
\end{tabular}
\\
&&\text{Table 2. Average and standard deviation of } \\
&&\rho (x,t_{s})\text{ and }\delta _{s}=\theta (x,t_{s})-\theta (x,0)\text{
for }\left| 2kx\right| \leq 16\pi \text{.}
\end{eqnarray*}
The improvements in the density and phase are evident as $\sigma (\rho
(x,t_{s}))$ and $\sigma (\delta _{s})$ converge monotonically to zero as $%
t_{s}$ increases. Furthermore, a\ focusing effect, like the one detailed
above, will occur near the times $t_{s}+t_{f}$ since the wave function at $%
t_{s}$ is nearly identical to $\phi (x,0)$.

Of course, the ability to perform an experiment on long time scales is
severely limited by transverse cooling considerations. An initial momentum
width of $\Delta p_{x}$ in the atomic beam washes out these coherent effects
in a time $t\lesssim \lambda M/\Delta p_{x}\sim (ku)^{-1}$, the Doppler
dephasing time. This condition does not present a problem for thick lens
focusing and lithography schemes\cite{3,14aa}. In order to see these thin
lens effects, the atom beam must be cooled or collimated near the recoil
limit for focusing ($ku\lesssim \omega _{k}$) or below the recoil limit for
Talbot self-imaging ($ku\ll \omega _{k}$). Recent experiments have conformed
with the focusing condition\cite{1,3,14bb}.

In summary, this letter has introduced the possibility of quasiperiodic atom
optical elements made from laser intensity gratings with incommensurate wave
vectors. The analytical results show that the atomic wave packet becomes a
quasiperiodic function, developing momentum components which are similarly
incommensurate and therefore regular, but not periodic. The atomic density
is a function of the time of flight from the diffraction grating. Atoms come
to semiclassical ''quasi''-focuses according to the depth and curvature of
the potential wells. The quasiperiodic density pattern can be used to create
a quasiperiodic surface for condensed matter studies when used for atomic
lithography. Furthermore, ultracold atoms will exhibit a quasi-self-imaging
of the wave function.

\acknowledgments

The authors would like to thank F. Nori, Q. Niu, G. Georgakis, and R. Merlin
for discussions regarding this work. This work is supported by the National
Science Foundation under Grant No. PHY-9414020, by the U.S. Army Research
Office under Grant No. DAAG55-97-0113, and by the University of Michigan
Rackham predoctoral fellowship.

\vspace{0.5cm}\centerline{\bf FIGURE CAPTIONS}\vspace{0.2cm}

Fig. 1. The atomic beam traverses the quasiperiodic potential formed by the
laser beams and is detected by light scattering or lithography after free
propagation.

Fig. 2. The optical potential in the interaction region for $A_{1}=5$ and $%
A_{2}=10$ and the corresponding atomic density at the quasi-focus, $%
z_{f}=v_{z}\left[ \omega _{k}(2A_{1}+A_{2})\right] ^{-1}$. Quasiperiods are
pronounced near $2kx\approx 2\pi ja_{s}\approx 2\pi jb_{s}\sqrt{2}$, where
peaks are labeled by $(ja_{s},jb_{s})$.

Fig. 3. Fourier amplitudes $\left| \rho _{m,n}(\omega _{k}t_{f})\right| $ at 
$q=2k(m+n/\sqrt{2})$ for the density in Fig. 2, as given by Eq. (\ref{7}).
Larger amplitudes are labeled by $(m,n)$. Inset: Full range of Fourier
components with significant amplitudes.

Fig. 4. Quasiperiodic Talbot Effect for $A_{1}=1$ and $A_{2}=1.$ We plot $%
\theta (x,t_{s})$, the phase of the atoms at the quasi-Talbot times $t_{s}$
for $s=2$ (-.-.-)$,4$ (....)$,6$ (- - -), versus $\theta (x,0)$, the initial
phase (---), where $\theta (x,0)=A_{1}\cos (2kx)+A_{2}\cos (\sqrt{2}kx)$.


\begin{references}
\bibitem{i}  For review articles, see {\it Atom Interferometry}, ed. by P.R.
Berman, Academic Press, San Diego (1997).

\bibitem{1}  V.P. Chebotayev {\it et al.}, J. Opt. Soc. Am. B {\bf 2}, 1791
(1985); O. Carnal and J. Mlynek, Phys. Rev. Lett. {\bf 66}, 2689 (1991); D.
Keith {\it et al.}, {\it ibid.} {\bf 66}, 2693 (1991); Rasel {\it et al.}, 
{\it ibid.} {\bf 75}, 2633 (1995)

\bibitem{2a}  Cahn {\it et al.}, Phys. Rev. Lett. {\bf 79}, 784 (1997)

\bibitem{2}  see B. Young, M. Kasevich, and S. Chu, in Ref. \cite{i}

\bibitem{3}  G. Timp {\it et al.}, Phys. Rev. Lett. {\bf 69}, 1636 (1992);
T. Sleator, V. Balykin, and J. Mlynek, Appl. Phys. B {\bf 54}, 375 (1992);
J.J. McClelland {\it et al.}, Science {\bf 262}, 877 (1993)

\bibitem{7}  H.F. Talbot, Philos. Mag. {\bf 9}, 401 (1836)

\bibitem{7aaa}  U. Janicke and M. Wilkens, J. Phys II (France) {\bf 4}, 1975
(1994); M.S. Chapman {\it et al.}, Phys. Rev. A {\bf 51}, R14 (1995)

\bibitem{7a}  N.P. Bigelow and M.G. Prentiss, Phys. Rev. Lett. {\bf 65}, 30
(1990); C.I. Westbrook {\it et al.}, {\it ibid.} {\bf 65}, 33 (1990)

\bibitem{7aa}  P. Verkerk {\it et al.}, Phys. Rev. Lett. {\bf 68}, 3861
(1992); P.S. Jessen {\it et al.}, {\it ibid.} {\bf 69}, 49 (1992)

\bibitem{8}  F.L. Moore {\it et al.}, Phys. Rev. Lett. {\bf 73}, 2974 (1994)

\bibitem{10}  R. Dum and M. Olshanii, Phys. Rev. Lett. {\bf 76}, 1788
(1996); M. Ben Dahan {\it et al.}, {\it ibid.} {\bf 76}, 4508 (1996)

\bibitem{11a}  P. J. Steinhardt and S. Ostlund, {\it The Physics of
Quasicrystals}, World Scientific, Singapore (1987); T.J. Fujiwara and T.
Ogawa, {\it Quasicrystals}, Springer-Verlag, Berlin (1993)

\bibitem{11b}  M. Ya. Azbel, Phys. Rev. Lett. {\bf 43}, 1954 (1979); B.
Simon, Adv. App. Math. {\bf 3}, 463 (1982); J.B. Sokoloff, Phys. Rep. {\bf %
126}, 1768 (1985)

\bibitem{14a}  M. Tanibayashi, J. Phys. Soc. Japan {\bf 61}, 3139 (1992)

\bibitem{12}  L. Guidoni {\it et al.}, Phys. Rev. Lett {\bf 79}, 3363 (1997)

\bibitem{13}  G.A. Georgakis, G. Sundaram, and Q. Niu, ''Quasi-Periodic
Lattices'' (unpublished, 1998)

\bibitem{14c}  The wave function $\phi $ is written in an interaction
picture for the $z$-dependent center-of-mass and spatially-independent
internal motions of the ground state.

\bibitem{14cc}  C. Henkel, J.-Y. Courtois, and A. Aspect, J. Phys. II
(France) {\bf 4}, 1955 (1994). The Raman-Nath condition restricts the
interaction time to $\tau \ll (\omega _{k}(V_{1}+V_{2}/2)/\hbar )^{-1/2}$. A
thin lens condition which should also be maintained here is $\tau \ll
(2\omega _{k}(V_{1}^{2}+V_{2}^{2}/2+\sqrt{2}V_{1}V_{2})/3\hbar ^{2})^{-1/3}$.

\bibitem{14e}  B. Dubetsky and P.R. Berman, Appl. Phys. B {\bf 59}, 147
(1994)

\bibitem{14aa}  K.K. Berggren {\it et al}., Science {\bf 269}, 1255, (1995);
S. Nowak, T. Pfau and J. Mlynek, Appl. Phys. B {\bf 63}, 203 (1996); K.S.
Johnson {\it et al.}, Appl. Phys. Lett. {\bf 69}, 2773 (1996)

\bibitem{14ee}  M. Kohmoto, B. Sutherland, and K. Iguchi, Phys. Rev. Lett. 
{\bf 58}, 2346 (1987); M.S. Vasconcelos, E.L. Albuquerque, and A.N. Mariz,
J. Phys.: Cond. Matt. {\bf 10}, 5839 (1998)

\bibitem{15}  The sequence is defined by the recurrence formulas, $%
a_{l}=a_{l-1}+$ $2b_{l-1}$ and $b_{l}=a_{l-1}+$ $b_{l-1}$ for $a_{1}=b_{1}=1$%
.

\bibitem{14bb}  J. Schmiedmayer {\it et al., }in Ref. \cite{i}, pp. 6-7
\end{references}
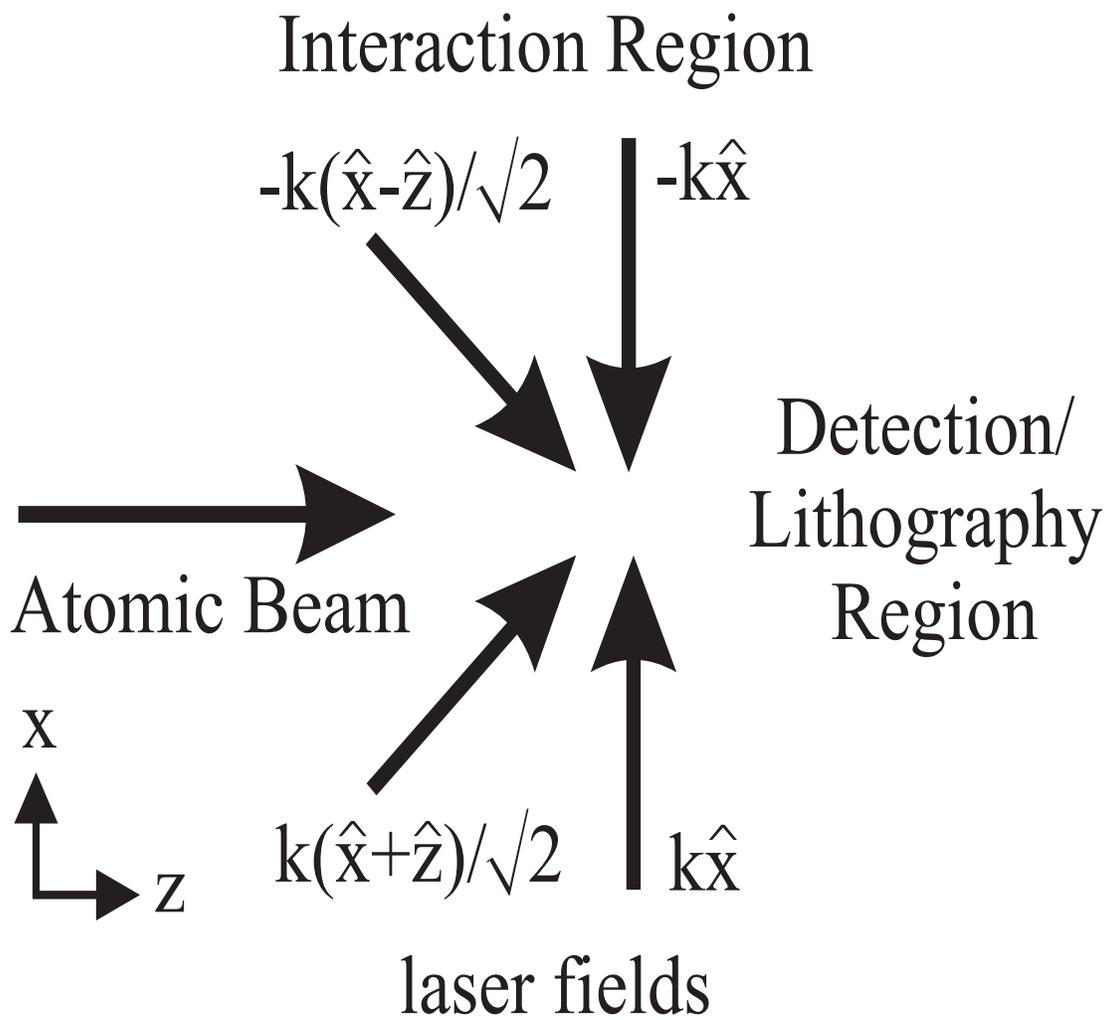
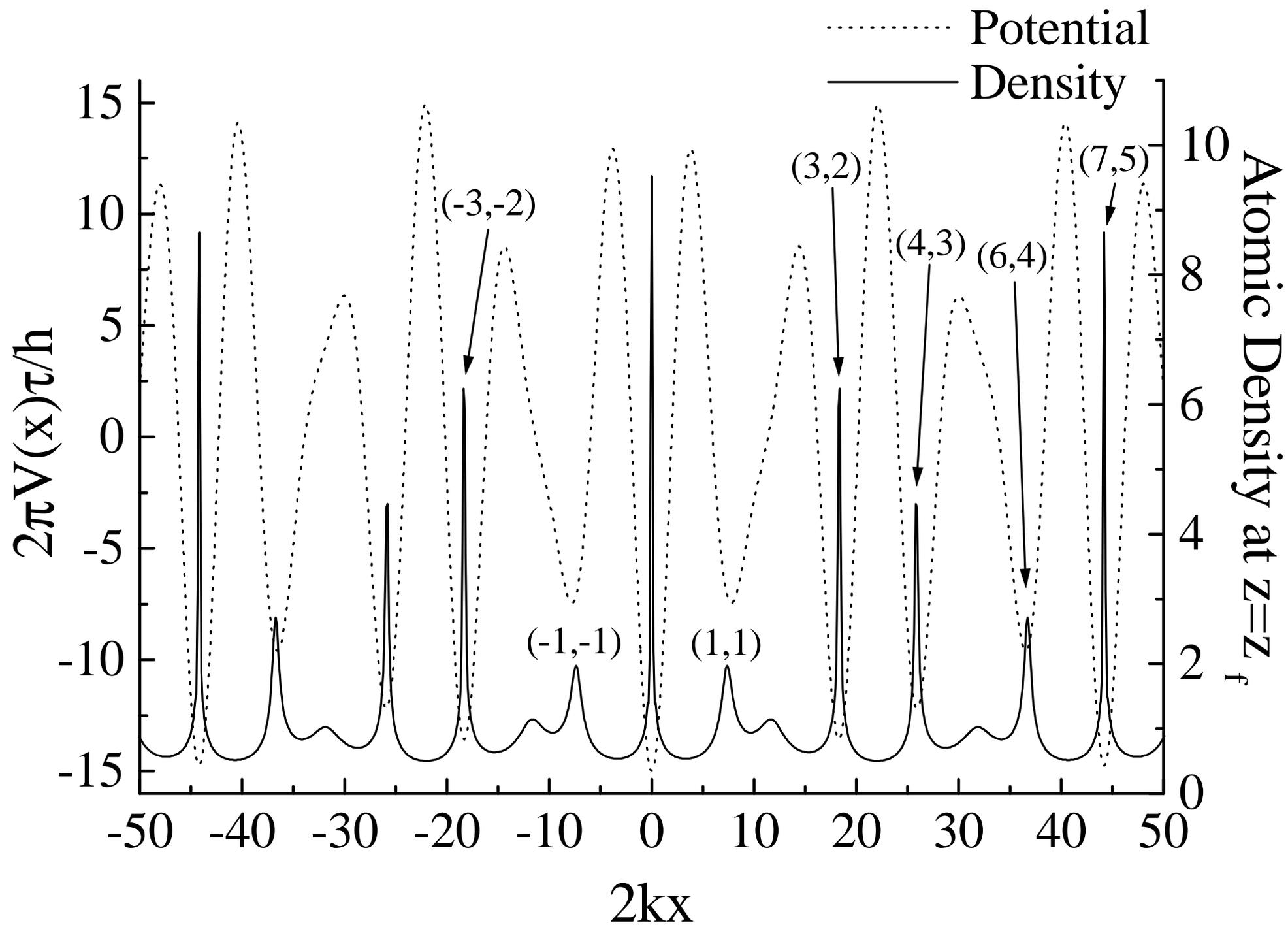
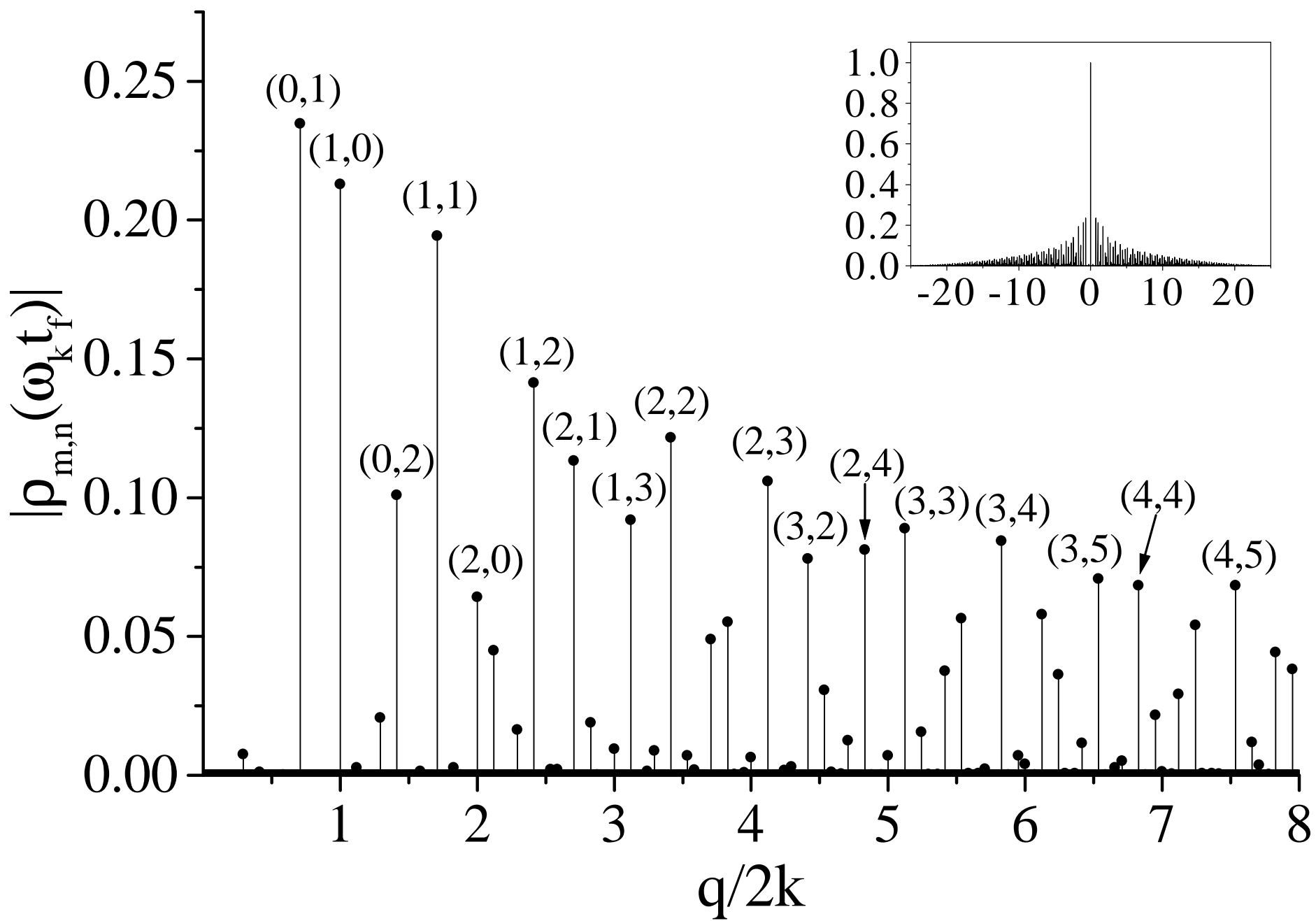
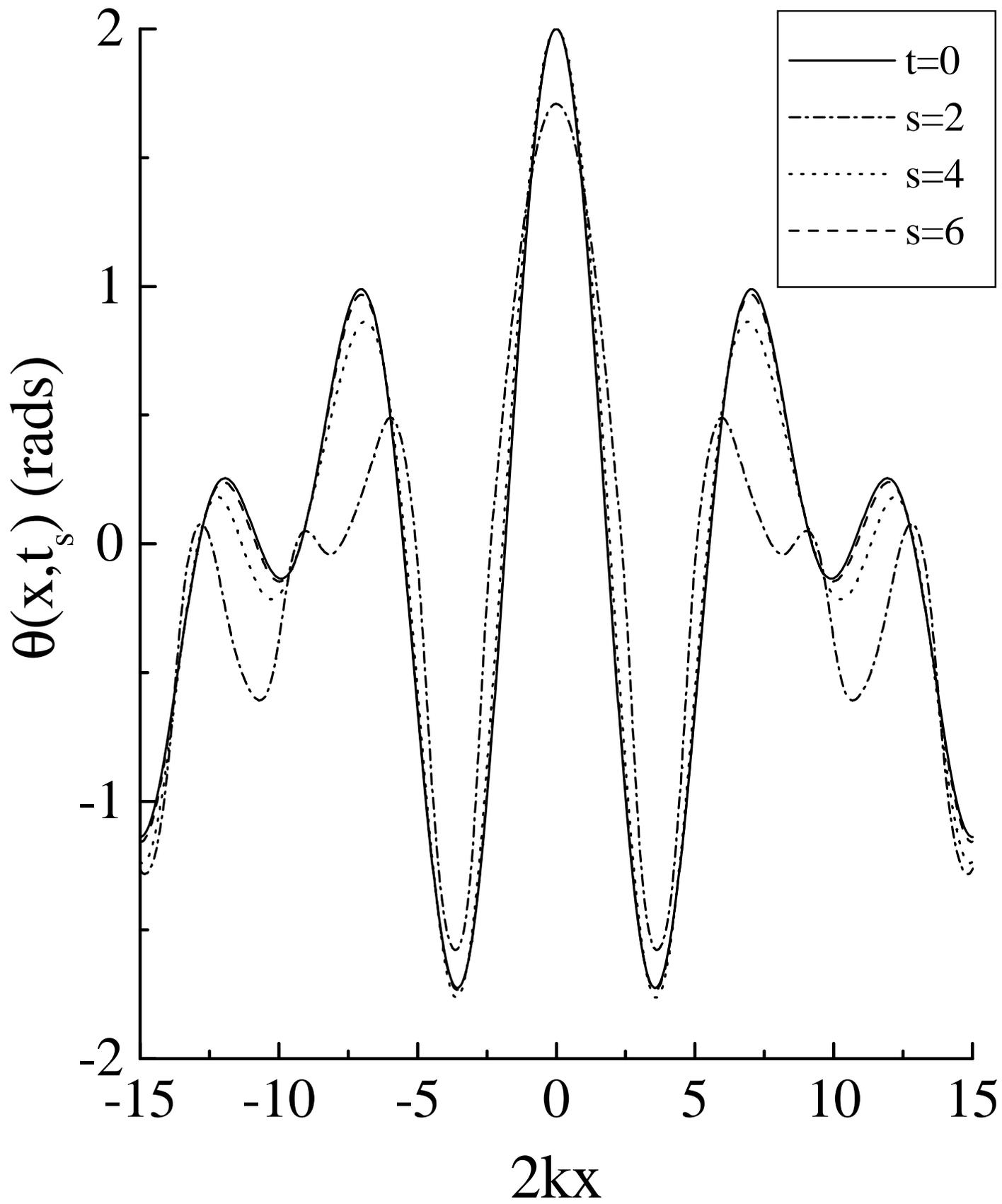
\end{document}